\documentclass[
    ,final            
  ]
  {aipproc}

\layoutstyle{8x11double}

\def\teff{T$_{\rm eff}$}

\begin{document}

\title{A Search for Binary Stars at Low Metallicity}


\author{David K. Lai}{
  address={UCO/Lick Observatory, University of California at Santa Cruz, Santa Cruz, CA 95064.}
}

\author{Sara Lucatello}{
  address={Osservatorio Astronomico di Padova, Vicolo dell'Osservatorio 5, 35122 Padua, Italy.}
}

\author{Michael Bolte}{
  address={UCO/Lick Observatory, University of California at Santa Cruz, Santa Cruz, CA 95064.}
}

\author{Debra A. Fischer}{
  address={Department of Physics and Astronomy, San Francisco State University, San Francisco, CA 94132}
}

\author{Jennifer A. Johnson}{
  address={Department of Astronomy, Ohio State University, Columbus, OH}
}

\begin{abstract}
We present initial results measuring the companion fraction of
metal-poor stars ([Fe/H]$<-$2.0). We are employing the Lick
Observatory planet-finding system to make high-precision Doppler
observations of these objects.  The binary fraction of metal-poor
stars provides important constraints on star formation in the early
Galaxy \citep{carney03}. Although it has been shown that a majority of
solar metallicity stars are in binaries, it is not clear if this is
the case for metal-poor stars. Is there a metallicity floor below
which binary systems do not form or become rare? To test this we are
determining binary fractions at metallicities below [Fe/H]$=-2.0$. Our
measurments are not as precise as the planet finders', but we are
still finding errors of only 50 to 300 m/s, depending on the
signal-to-noise of a spectrum and stellar atmosphere of the star. At
this precision we can be much more complete than previous studies in
our search for stellar companions.
\end{abstract}

\maketitle


\section{Introduction}
The observational study of mutliple star systems has been carried out
extensively by previous groups (e.g., \citep{dm,carneylatham87}).  The general
results have shown that among solar neighborhood F and G stars, there
is a multiple star system fraction of approximately
65\%. \citet{carney05} recently discussed the incidence of binarity
among metal-poor stars in the context of a deficiency of metal-poor
binaries on galactic retrograde orbits. \citet{lucatello06} also
looked at binary frequency in metal-poor stars, but only in the subset
of carbon-enhanced, s-process rich objects, finding a result that is
consistent with 100\% binarity.

As discussed above, much work has been done in determining the
fraction of stars from solar down to low metallicities.  But until
only recently, have large enough number of metal-poor stars been
discovered to begin exploring this regime systematically. What we are
doing in this current study is extending this search into the very low
metallicity with a new method.  We are using the iodine cell
radial velocity monitoring techniques developed for the planet finding
group at Lick Observatory to get highly precise radial
velocities for a number of very metal-poor stars.

\section{Method}

Our observational setup is very similar to the Lick Observatory
Iondine Planet Search group \citep{fischer03}. We use a heated iodine
cell placed into the light path of the Hamilton spectrometer to create
a series of reference absorption lines. This enables us to obtain very
precise radial velocities measuremnts, irrespective of the true
zero-point radial velocity \citep{butler96}. 

Although the observational setup and tools are the same as the planet
search team, we face two unique challenges. The first is that our
stars are generally faint, $V=8$ and dimmer. This means we have much
lower SNR spectra. The second challenge is that very metal-poor stars
have fewer absorption lines, and therefore less radial velocity
information than the near solar-metallicity stars of the planet
searchers. 

Nevertheless, we have begun devoloping a technique that still gives
very precise velocites and errors ranging from only 50 to 300 m/s.

\section{Data}

In this ongoing project we have time baselines ranging from days up to
almost 3 years. We are also observing both C-rich and C-normal
metal-poor stars.

Our data so far is comprised of the following.
\begin{itemize}
\item
43 objects with 3 or more observations.
\item
36 objects with 4 or more observations.
\item
33 objects with 5 or more observations.
\end{itemize}

\section{Preliminary Results}

\begin{figure}
  \includegraphics[height=.45\textheight]{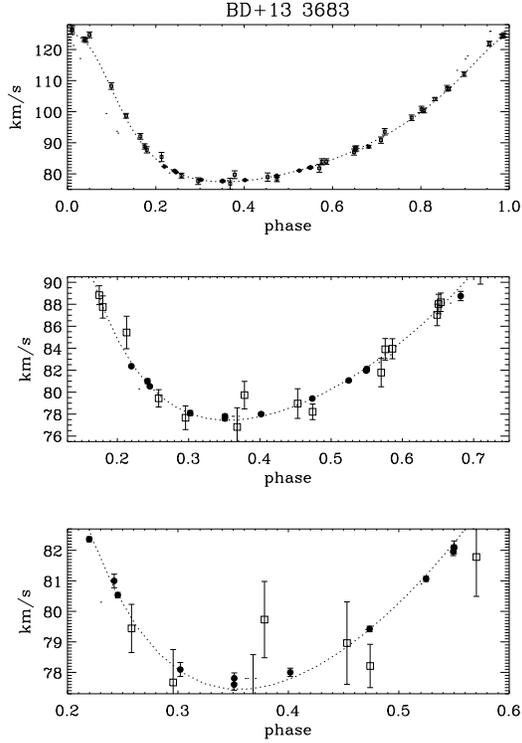}
  \caption{The known binary star BD+13-3683 ([Fe/H]$\sim$2.0). We show
  the data and orbital solution from \citet{carney03} as the
  square points and dotted line. Our measurements are plotted as the
  black dots, and the error bars represent 2 times the actual error.
  \label{proofofconcept}}
\end{figure}

\begin{figure}
  \includegraphics[height=.2\textheight]{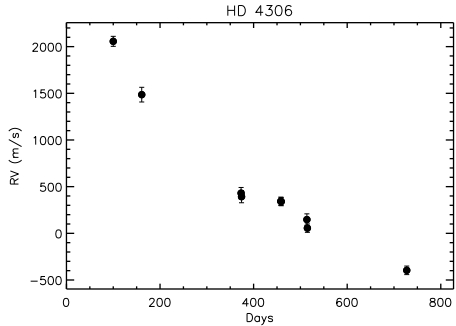}
  \caption{The radial velocity
measurements for HD 4306.
  \label{hd4306}}
\end{figure}

\begin{figure}
  \includegraphics[height=.2\textheight]{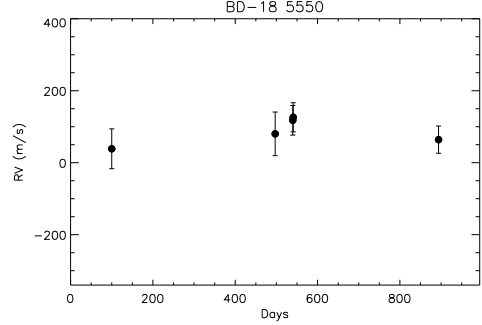}
  \caption{An example of a star
where we find no radial velocity variation.
  \label{bd-18}}
\end{figure}

\begin{figure}
  \includegraphics[height=.2\textheight]{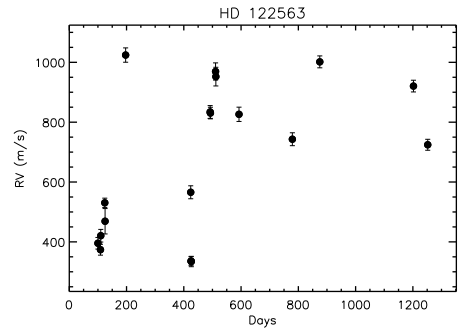}
  \caption{The measurements for HD 122563. At its current stage of
    stellar evolution (\teff{}$\sim$4600 K) it is unclear if this is a
    signature of a companion or an intrinsic velocity jitter.
  \label{hd122563}}
\end{figure}

Our prelimanary results have been very encouraging. In figure
\ref{proofofconcept} we show a test of our method. Figures
\ref{hd4306} to \ref{hd122563} give examples of three results: a
detection of a companion, a non-detection, and the detection of either a
companion or a intrinsic velocity jitter.

\begin{figure}
  \includegraphics[height=.35\textheight]{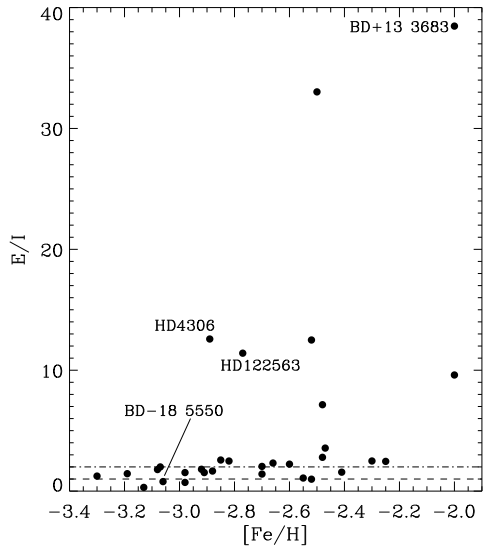}
  \caption{For each object, the rms of the velocity measurements, E,
  divided by the average radial velocity error for that object,
  I. The higher the value of E/I, the more likely it is a true
  radial velocity variable.
  \label{EI}}
\end{figure}

Overall we have completed the first-pass analysis of 31 of our
objects. The [Fe/H] distribution and E/I values are shown in figure
\ref{EI}. We find that 55\% of our targets have detected radial
velocity variation (defined here as E/I $>$2).

If we eliminate stars with \teff{}$<4700$K to minimize the potential
effect of intrinsic stellar jitter, we are left with 26 stars, 12 of
which have clear radial velocity variations. If these variations are
assumed to be caused by a companion, the we find a spectrosopic
binary fraction of 46\%.



\begin{theacknowledgments}
We would like to acknowledge the National Science Foundation for their support
under grant AST-060770.

\end{theacknowledgments}



\bibliographystyle{aipproc}   

\bibliography{all.bib}

\IfFileExists{\jobname.bbl}{}
 {\typeout{}
  \typeout{******************************************}
  \typeout{** Please run "bibtex \jobname" to optain}
  \typeout{** the bibliography and then re-run LaTeX}
  \typeout{** twice to fix the references!}
  \typeout{******************************************}
  \typeout{}
 }

\end{document}